\documentclass[aps,prd,reprint]{revtex4-1}

\usepackage{graphicx}
\usepackage[utf8]{inputenc} 
\usepackage[T1]{fontenc}    
\usepackage{hyperref}       
\usepackage{url}            
\usepackage{booktabs}       
\usepackage{amsfonts}       
\usepackage{nicefrac}       
\usepackage{microtype}      

\newcommand{\GeV}{GeV/c$^2$}
\newcommand{\MJD}{\textsc{Majorana Demonstrator}}

\begin{document}

\title{Direct Detection Limits on Heavy Dark Matter}
\author{Michael Clark}
\email{clark632@purdue.edu}
\author{Amanda Depoian}
\author{Bahaa Elshimy}
\author{Abigail Kopec}
\author{Rafael F. Lang}
\author{Shengchao Li}
\email{li4006@purdue.edu}
\author{Juehang Qin}
\affiliation{Department of Physics and Astronomy, Purdue University, West Lafayette, IN 47907, USA}

\begin{abstract}
Multiply-interacting massive particles (MIMPs) are heavy ($>10^{10}$~\GeV) dark matter particles that interact strongly with regular matter, but may have evaded detection due to the low number density required to make up the local dark matter halo.  These particles could leave track-like signatures in current experiments, similar to lightly-ionizing particles.  We show that previously calculated limits from the \MJD~on the flux of lightly-ionizing particles can be used to exclude MIMP dark matter parameter space up to a mass of $10^{15}$~\GeV.  We also calculate limits from the standard XENON1T analysis in this high-mass regime, properly taking into account flux limitations and multi-scatter effects. Finally, we show that a dedicated MIMP analysis using the XENON1T dark matter search could probe unexplored parameter space up to masses of $10^{18}$~\GeV.
\end{abstract}

\maketitle

\section{Introduction}

There has been some interest in particle dark matter with masses close to the Planck mass ($10^{19}$~\GeV)~\cite{Bramante:2018qbc,Bramante:2018tos,Bramante:2019yss,Lehmann:2019zgt,Cappiello:2020lbk,Meissner:2018cay}, though direct searches are lacking.  
Such heavy dark matter particles can arise from supersymmetric theories~\cite{Raby:1997pb} and be produced in the early universe~\cite{Chung:1998zb,Kolb:2017jvz,Bhoonah:2020oov}. Dark matter could also consist of composite objects at this mass, such as primordial black holes~\cite{Lehmann:2019zgt,Hooper:2019gtx} or other composite particles~\cite{Coskuner:2019,Ponton:2019}.
However, the sensitivity of direct dark matter detection experiments for a given exposure will decrease linearly as the mass of the dark matter particles increases.
This is due to the fact that with a constant local dark matter density~\cite{Read:2014qva}, more massive individual particles lead to a lower number density.
As a result, detection sensitivity to high mass dark matter particles will be limited to high interaction cross-sections.

As the dark matter-nucleon cross-section increases, the likelihood will also increase that a particle interacts more than once as it passes through the experiment.
Events with scattering multiplicity larger than one are typically not considered as signal candidate events searches due to the typically expected cross-sections being low.  
Searching for such multiple-scatter events will allow dark matter detectors to access a wider parameter space, both at higher masses and higher cross-sections.  
Particles that interact in this manner can be called Multiply-Interacting Massive Particles, or MIMPs~\cite{Bramante:2018qbc}.  
At masses much higher than the mass of a nucleus, MIMP-nucleus scattering will result in a negligible change in the MIMP's momentum due to its large kinetic energy.
Therefore, MIMPs will travel in a straight path through matter, despite any number of scatters.

Searches for lightly ionizing particles, such as those performed by CDMS-II~\cite{Agnese:2014vxh} or the \MJD~\cite{Alvis:2018yte}, are specifically targeting such multiple scatter events.
These searches did not find any excess events in their region of interest.
This provides an opportunity to set limits on MIMP dark matter with experiments that have already been carried out, by reinterpreting the results of these searches. 
TEXONO~\cite{Singh:2018von} has also recently performed a search for lightly-ionizing particles with sensitivity to much lower charges, though only considering single scatter events.
In this study, we determined the signature of a multiple-scatter dark matter event in both the \MJD~and the XENON1T experiments, using the local dark matter halo properties and the geometry of the detectors.
These experiments were chosen since they hold the most sensitive results for lightly-ionizing particles~\cite{Alvis:2018yte} and WIMP dark matter~\cite{Aprile:2018dbl} respectively.
We used these multiple-scatter signatures to set limits on the interaction cross-section and mass of dark matter particles up to $10^{18}$~\GeV.

\section{Procedure}

\begin{figure}
  \centering
  \includegraphics[width=\columnwidth]{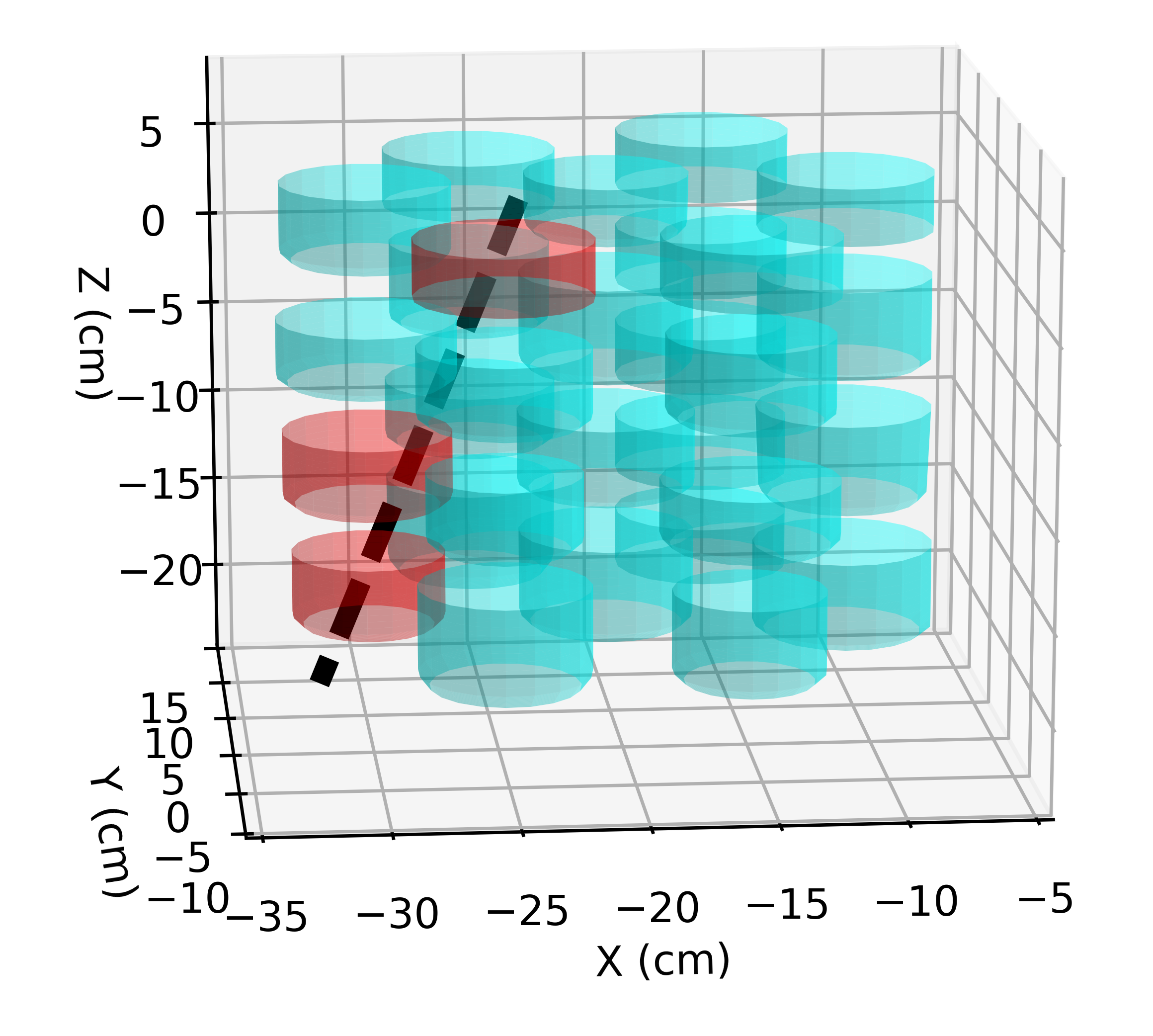}
  \caption{An example of a simulated MIMP track through one of the \MJD~detector modules.  This track has passed through three of the thirty detector units (highlighted in dark red).  Such simulated tracks are used to determine the energy deposited by MIMPs entering the detector.}
  \label{fig:simulation}
\end{figure}

The \MJD~is a low-background rare-event search located at the Sanford Underground Research Facility, employing a total of 57 point-contact germanium detectors to search for rare events, primarily from neutrinoless double-beta decay~\cite{Abgrall:2013rze}.  
The germanium detectors are organized into towers of three to five units, and fourteen towers are assembled into two hexagonal modules.  Figure~\ref{fig:simulation} shows the general setup of one of the two detector modules.  
These modules are enclosed within a two-inch thick plastic scintillator muon veto to monitor for through-going muons~\cite{Alvis:2018pne}.

The \MJD~was used to search for lightly ionizing particles~\cite{Alvis:2018yte} by targeting coincident energy depositions between more than one detector unit within its detector modules, as lightly-ionizing particles will pass through the detector depositing small amounts of energy along their path. 
Zero candidate events were observed with a detector unit multiplicity of 4-6, in an exposure of 285 days of data taking~\cite{Alvis:2018yte}.

An equivalent signature would be seen from heavy particles ($>10^{10}$~\GeV) with a high MIMP-nucleus cross-section, that would not lose significant momentum or energy even when colliding with many atoms along their path. 
Assuming canonical elastic scattering off nuclei~\cite{Baudis:2012ig}, with $\theta$ being the scattering angle in the center-of-mass frame, $v$ the velocity of the MIMP, $m_N$ the mass of the nucleus, $M$ the mass of the MIMP, and $\mu$ the MIMP-nucleus reduced mass, the energy $E_R$ transferred to the recoiling nucleus at each collision can be written as:
\begin{equation}
    E_R = \frac{\mu^2v^2}{m_N}(1-\cos\theta) \approx m_Nv^2(1-\cos\theta) \ll \frac{1}{2}Mv^2,
    \label{eq:energy}
\end{equation}
since $\mu\approx m_N$ if $M \gg m_N$.  In the lab frame, the maximum deflection angle of the MIMP from each collision can be found using elastic scattering kinematics~\cite{Griffiths:2008zz}:
\begin{equation}
    \sin\theta = m_N/M \approx 0.
    \label{eq:angle}
\end{equation}
The combination of Eqs.~\ref{eq:energy} and \ref{eq:angle} show that a MIMP will travel in a straight line with a constant velocity through any dark matter detector with $m_N \ll M$, even if its cross-section is high.
The coincidence window used by the \MJD~(4~$\mu$s)~\cite{Alvis:2018yte} is long enough such that a dark matter particle moving at $\mathcal{O}(100)$~km/s, as expected from the local WIMP halo parameters~\cite{jelle_aalbers_2019_3345959}, would cross the full detector within one coincidence window, and would be observed within the dataset.

We simulated MIMP events to determine the sensitivity of the \MJD~to MIMP dark matter. 
Standard WIMP halo model values~\cite{jelle_aalbers_2019_3345959} were assumed for the local MIMP density and velocity distribution.
The geometry of the \MJD~detector array and exposure times for each data taking run were taken from \cite{Alvis:2019sil}.
The full detector modules were assumed to be operational for each data taking run that they were employed.
It was confirmed through simulation that one detector not being operational during a run had a negligible impact on the resulting overall limit.
Straight tracks were simulated with an isotropic distribution through the detector, with an additional acceptance loss at high cross-section due to attenuation from the earth overburden~\cite{Bramante:2018qbc}.
These tracks were used to determine the detector unit multiplicity, and the path length through each unit. 
An example simulated track passing through one module of the \MJD~is shown in Figure~\ref{fig:simulation}, with the displayed event having triggered three of the thirty detector units in the module.  

For a given MIMP mass, a number of MIMPs are simulated using Poisson statistics according to the expected flux of dark matter during the exposure.  
Then, the path travelled by each MIMP through the detector is simulated, and a number of interactions within each detector unit is simulated using the MIMP-nucleon cross-section.
For each collision, the amount of energy deposited is simulated according to the kinematics between the MIMP mass and nuclear mass as in Eq.~\ref{eq:energy}, and then converted to an effective energy deposition using the nuclear quenching factor from the Lindhard model~\cite{Lindhard:1963}, which has been shown to agree well with observations in germanium down to low energies~\cite{Scholz:2016qos}.  
A detector unit is considered triggered if the deposited energy exceeds the per-unit energy threshold, which varies per detector unit between 0.8 and 2~keV~\cite{Alvis:2018yte}.  
As the crossing time of a particle is shorter than the coincidence window, all interactions that occur within a detector unit will contribute to the total amount of deposited energy for that event.

The \MJD~employs a plastic scintillator muon veto around the detector arrays.  
In the lightly-ionizing particle search, any event within one second of an energy deposition greater than 1~MeV in the muon veto was removed.  
If the MIMP-nucleon cross-section is sufficiently high, then a MIMP will deposit enough energy within the muon veto to cause a trigger, and such events would have been removed.  
This means that the search considered here did not have sensitivity to particles that deposit an energy above 1~MeV in the muon veto, which corresponds to particles with cross-sections $>10^{-28}$~cm$^2$.

In addition to the \MJD, limits for the currently leading WIMP detection experiment XENON1T~\cite{Aprile:2018dbl} were also simulated using the same procedure, but with the XENON1T geometry.  
This was done both to verify the limits set by this detector at such high masses, which are not usually considered in a WIMP search, as well as to determine what a dedicated multiple scatter search could yield in terms of sensitivity in this parameter range.  
The simulations were carried out using an exposure time of 279 days~\cite{Aprile:2018dbl}.
Since XENON1T is a monolithic liquid xenon detector instead of a modular germanium detector, the detection strategy would need to be adjusted for events with high multiplicity.  
With the ability of a xenon time projection chamber to resolve individual interaction vertices, MIMP track-like events could be searched-for by looking for events with high multiplicity, oriented in a straight line.  
Since neutron and gamma interactions can also interact multiple times within the detector, one could select for events with high multiplicities and linear topology to obtain a purer MIMP sample at the expense of sensitivity.

\section{Results and Discussion}

\begin{figure}[!h]
  \centering
  \includegraphics[width=\columnwidth]{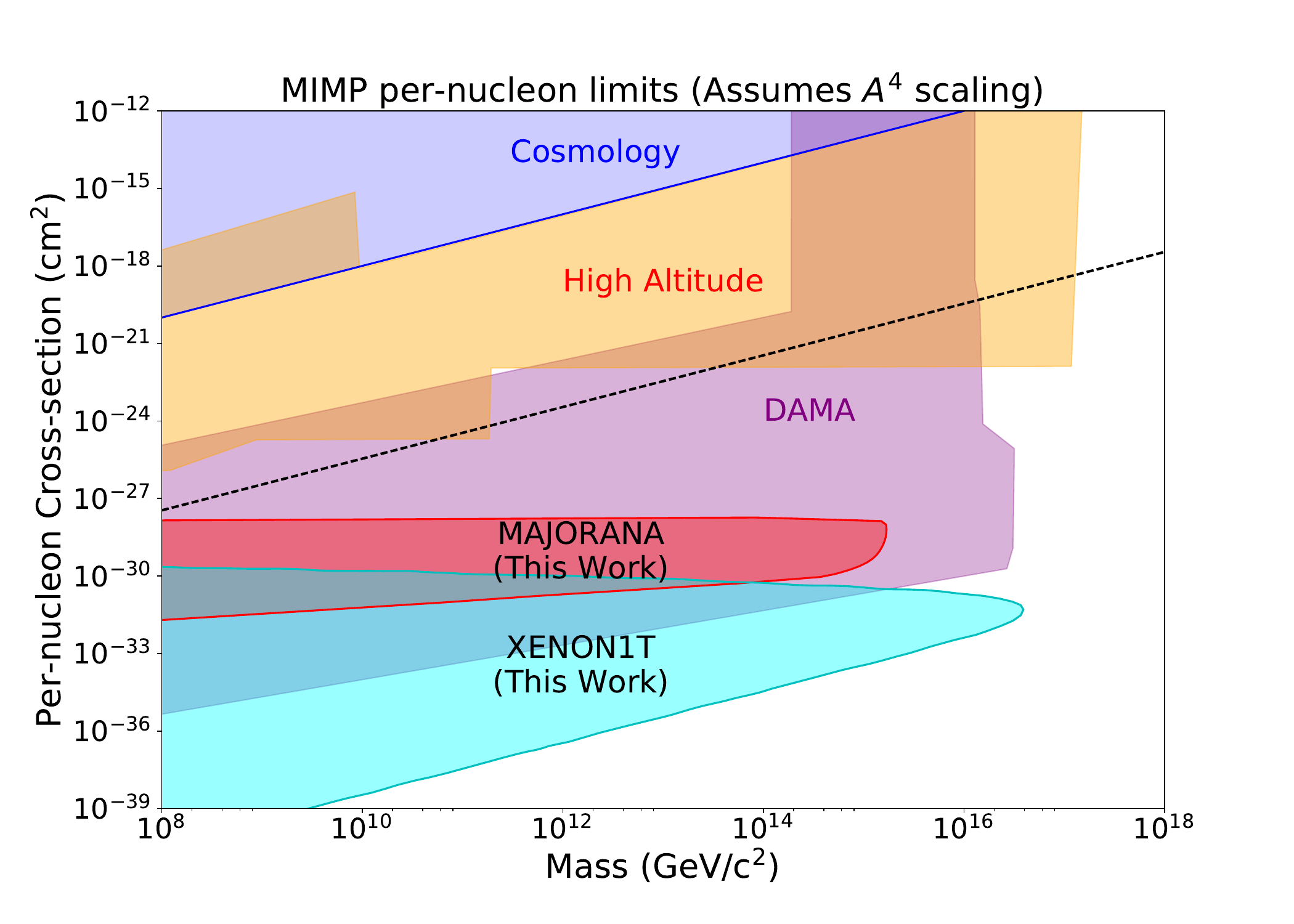}
  \caption{Excluded regions in the parameter space of MIMP-nucleon cross-section vs. MIMP mass from various sources~\cite{Wandelt:2000ad,Mack:2007xj}, as well as the limits calculated here from the \MJD~(red) and XENON1T~(cyan). The dashed black line shows the maximum sensitivity for underground detectors due to the overburden of the Earth's crust. $A^4$ scaling is assumed due to coherent interactions with the nucleus.}
  \label{fig:limits}
\end{figure}

Figure~\ref{fig:limits} shows the results obtained from our simulations, as well as previous results from astrophysical constraints~\cite{Cyburt:2002uw}, high altitude particle detectors~\cite{McGuire:1994pq,Price:1975bm} and the DAMA collaboration~\cite{Bernabei:1999ui} corrected for saturated overburden scattering~\cite{Bramante:2018qbc}.
At MIMP-nucleon cross-sections above $1.3\times10^{-20} \mathrm{cm}^2$, the MIMP will scatter with every nucleus in the Earth's crust above the detector, and thus increasing the cross-section above this limit will no longer increase the stopping power of the Earth~\cite{Bramante:2018qbc}.
These limits are shown in the MIMP-nucleon cross-section vs. MIMP mass parameter space to allow comparison between different detector materials.  
Particles with cross-sections above the black-dashed line will no longer reach the detector through the Earth overburden.
We have chosen to display the results starting at MIMP masses above $10^{8}$~\GeV for clarity in the interested region, but these limits could be extended to lower masses.

\begin{figure}
  \centering
  \includegraphics[width=\columnwidth]{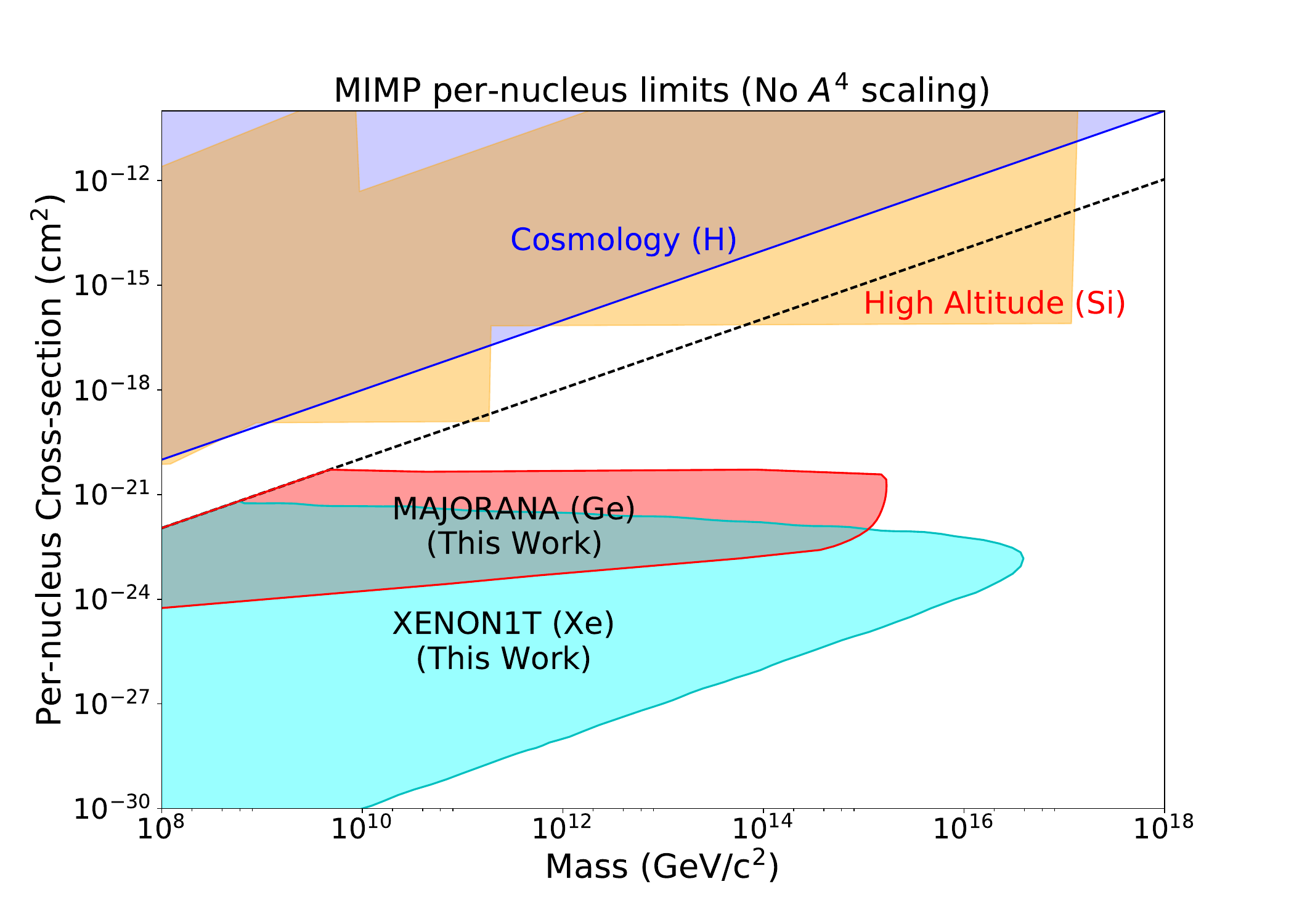}
  \caption{Excluded regions in the parameter space of MIMP-nucleus cross-section vs. MIMP mass from various sources~\cite{Wandelt:2000ad,Mack:2007xj}, as well as the limits calculated here from the \MJD~(red) and XENON1T~(cyan). The corresponding nucleus used to correct for the $A^4$ scaling is shown in the labels of each region.}
  \label{fig:nuclimits}
\end{figure}

Limits from astrophysical constraints are primarily derived from scattering off of hydrogen, whereas the particle detectors use different nuclei that benefit from an $\propto A^4$ increase in the scattering cross-section due to coherence~\cite{Goodman:1984dc,Digman:2019wdm}.
Recently it was pointed out that this $\propto A^4$ scaling does not apply for point-like particle dark matter above cross-sections of $10^{-31}$cm$^2$~\cite{Digman:2019wdm}.
At cross-sections comparable or larger than the geometric size of the nucleus, the form factor of the interaction no longer follows the Born approximation, and thus the model-independent coherence effect no longer holds.
This is true for both contact interactions with heavy mediators, and longer-range interactions with light mediators, though interactions with light mediators can maintain coherence at higher cross-sections up to $10^{-25}$cm$^2$~\cite{Digman:2019wdm}.
However, a full treatment of coherence of dark matter-nucleus scattering at such large cross-sections is currently lacking.
Without correcting for this coherent enhancement in each different nucleus, the results from different experiments cannot be directly compared without assuming a certain model.  
Therefore, to compare with other experiments and with previously published limits we have chosen to display the results in two different parameter spaces.
In addition to the per-nucleon scaling in Figure~\ref{fig:limits}, we also show the limits in reference to the per-nucleus cross-section in Figure~\ref{fig:nuclimits}. This cross-section is relevant for composite dark matter candidates, with a cross-section equal to the physical size of the composite particle~\cite{Cappiello:2020lbk}.  To put previously published limits into this space, the cross-section is corrected to remove $A^4$ for the corresponding material.
The dashed line from overburden attenuation is also scaled according to the average atomic mass of the materials in the Earth's crust~\cite{Bramante:2018qbc}.
The uncertainty in the scaling of the cross-sections highlights the value in carrying out experiments with different target materials to support and corroborate with existing limits.

The region excluded by XENON1T shown in Figure~\ref{fig:limits} differs from regions claimed previously~\cite{Kavanagh:2017cru}, since our treatment takes into account that the normal WIMP search strategy does not consider events with multiplicity greater than one.  
For the XENON1T detector, using the density of liquid xenon, the mean free path of a MIMP becomes equal to the size of the detector ($\sim$1~m) once the cross-section exceeds $10^{-32}$~cm$^2$. 
This leads to the result that the XENON1T WIMP analysis loses sensitivity at high cross-sections.

\begin{figure}
  \centering
  \includegraphics[width=\columnwidth]{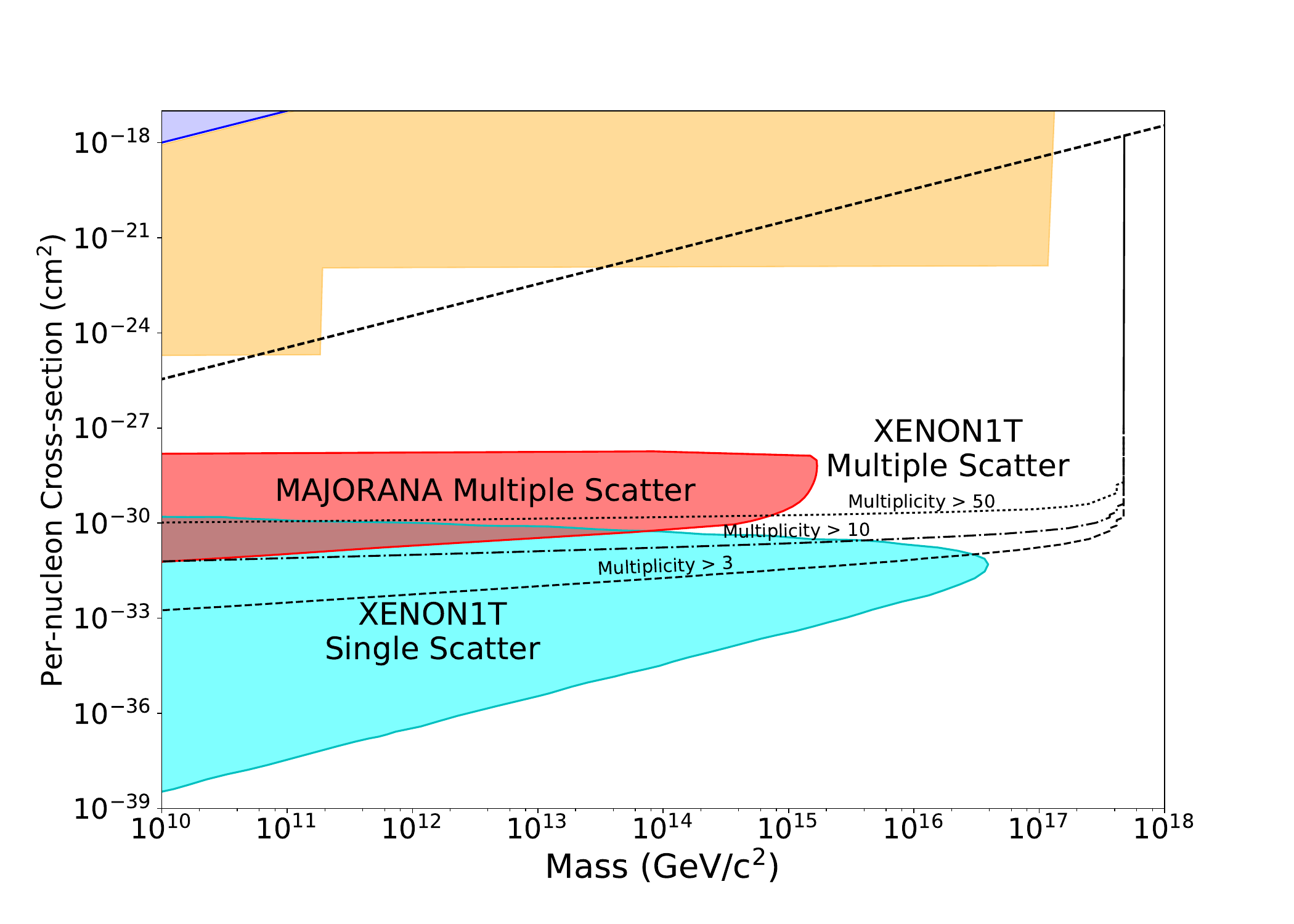}
  \caption{The possible sensitivity curves for a dedicated MIMP analysis of the XENON1T dark matter search data (Assuming $A^4$ scaling as in Fig.~\ref{fig:limits}), considering events with scattering multiplicities greater than 3 (dashed), greater than 10 (dash-dot), and greater than 50 (dotted). }
  \label{fig:sens}
\end{figure}

The region excluded by the \MJD~reaches to higher cross-sections than probed by the XENON1T experiment, due to the dedicated search for tracks.
The upper limit of this region is determined by the muon veto, which would have been triggered by MIMPs with higher cross-section.

Figure~\ref{fig:sens} shows the exclusion regions calculated in this study, with additional lines to show the possible sensitivity of a dedicated MIMP analysis of the XENON1T exposure. 
Each line shows the lower limit of sensitivity for a search for multiple interaction events with a certain minimum multiplicity, assuming no background events.
Regardless of the multiplicity considered, there is a strong bound on the sensitivity near $10^{18}$~\GeV where the XENON1T experiment becomes flux limited.
At this mass, the local particle density of dark matter becomes low enough that less than one particle is expected to cross the detector during the entire exposure of the experiment.
This analysis would not have the same constraint on the upper bound of the region as the \MJD~due to the XENON1T muon veto being a water Cherenkov detector~\cite{Aprile:2014zvw}. 
The MIMPs are not charged, and thus would not trigger the XENON1T muon veto.  
This would allow such a study to reach much higher cross-sections than previously reached by dark matter direct detection experiments~\cite{Kavanagh:2017cru}.

\section{Conclusion}

Multiply-interacting, heavy dark matter particles are a new region of interest in dark matter direct detection.  
These types of particles are motivated by theory, but the relevant parameter space is difficult to access and requires dedicated analyses.
By reinterpreting the lightly ionizing particle search of the \MJD~as limits on strongly interacting dark matter, we ruled out previously un-probed parameter space.  
In addition, the limits that have been set by XENON1T were verified in this high mass region of parameter space.  
A dedicated XENON1T MIMP analysis using existing data would be able to push this sensitivity region much higher.

\section{Acknowledgements}

We are grateful to Ralph Massarczyk from the \textsc{Majorana} Collaboration for helpful discussions and information on the \MJD. 
We are also grateful to Nirmal Raj, Ibles Samblas, and Shawn Westerdale for useful discussions.
This work was carried out thanks to support from the National Science Foundation through grant no. PHY-1719271.

\bibliographystyle{unsrt}  
\bibliography{references}

\end{document}